\documentclass{article}
\usepackage[utf8]{inputenc}
\usepackage{graphicx}
\usepackage{amsmath}
\usepackage{amssymb}
\usepackage{wrapfig}
\usepackage{float}
\usepackage[sorting=none]{biblatex}
\bibliography{final.bib}
\usepackage{hyperref}
\usepackage{booktabs}
\usepackage{siunitx}
\usepackage{amssymb}
\usepackage{bm}
\usepackage{tcolorbox}
\usepackage[legalpaper, margin=1in]{geometry}
\usepackage{subcaption}
\usepackage{newunicodechar}

\newcommand{\RR}{\mathbb{R}}

\providecommand{\keywords}[1]
{
  \small	
  \textbf{\textit{Keywords---}} #1
}

\usepackage{newunicodechar}
\DeclareRobustCommand{\okina}{%
  \raisebox{\dimexpr\fontcharht\font`A-\height}{%
    \scalebox{0.8}{`}%
  }%
}
\newunicodechar{ʻ}{\okina}

\title{All Thresholds Barred: Direct Estimation of Call Density in Bioacoustic Data}

\author{Amanda K. Navine$^{1\dagger}$, Tom Denton$^{2\dagger}$, Matthew J. Weldy$^{3}$, Patrick J. Hart$^{1}$\\
    \small $^{1}$Listening Observatory for Hawaiian Ecosystems, University of Hawai‘i at Hilo, Hilo, HI, USA\\
    \small $^{2}$Google Research, San Francisco, CA, USA\\
    \small $^{3}$Department of Forest Ecosystems and Society, Oregon State University, Corvallis, OR, USA\\
    \small $^{\dagger}$These authors share first authorship }

\begin{document}
\maketitle
\begin{abstract}
Passive acoustic monitoring (PAM) studies generate thousands of hours of audio, which may be used to monitor specific animal populations, conduct broad biodiversity surveys, detect threats such as poachers, and more. Machine learning classifiers for species identification are increasingly being used to process the vast amount of audio generated by bioacoustic surveys, expediting analysis and increasing the utility of PAM as a management tool. In common practice, a threshold is applied to classifier output scores, and scores above the threshold are aggregated into a detection count. The choice of threshold produces biased counts of vocalizations, which are subject to false positive/negative rates that may vary across subsets of the dataset. In this work, we advocate for directly estimating call \emph{density}: The proportion of detection windows containing the target vocalization, regardless of classifier score. Our approach targets a desirable ecological estimator and provides a more rigorous grounding for identifying the core problems caused by distribution shifts --- when the defining characteristics of the data distribution change --- and designing strategies to mitigate them. We propose a validation scheme for estimating call density in a body of data and obtain, through Bayesian reasoning, probability distributions of confidence scores for both the positive and negative classes. We use these distributions to predict site-level densities, which may be subject to distribution shifts. We test our proposed methods on a real-world study of Hawaiian birds and provide simulation results leveraging existing fully annotated datasets, demonstrating robustness to variations in call density and classifier model quality.

\keywords{Bioacoustics, Machine Learning, Wildlife Monitoring, Bayesian Modeling, Predictive Reasoning}
\end{abstract}

\section{Introduction}

Slowing the alarming pace of global biodiversity loss will require the development of tools and protocols for effective wildlife population monitoring and management  \cite{Butchart2010}. Understanding population responses to threats and conservation actions is critical in developing successful conservation strategies \cite{Nichols2006}. Passive acoustic monitoring (PAM) using automated recording units has become an increasingly used technique in wildlife management as it can provide a non-invasive and cost-effective approach to collecting data on sound-producing species, including those with cryptic behaviors or in difficult-to-survey habitats \cite{Sugai2019}. However, PAM can generate large quantities of acoustic data, necessitating machine and deep learning algorithms to detect species of interest semi-automatically  \cite{Tuia2022}. These computational tools now at our disposal present a promising opportunity, but inferring accurate biological and ecological significance from the results remains a challenge \cite{Gibb2018}. 

In many applications, a classifier score threshold is selected, and classifier outputs are reduced to binary detections \cite{Knight2017,Cole2022}, which creates false positives and false negatives that need to be accounted for in downstream work \cite{Miller2015,Chambert2018,Clement2022}. Counts of these detections are a tempting proxy for activity levels, but the ambiguity introduced by mis-detections makes such interpretation difficult. As a result, binary detection counts are often further aggregated to binary indicators of detection-nondetection by validating high-scoring examples. This approach can mitigate the risk of false positives, but at the risk of higher false negative rates \cite{Knight2017}.

However, the underlying \emph{call density} $P(\oplus)$ is a compelling target. When greater than zero, call density is an occupancy indicator. After establishing occupancy ($P(\oplus)>0$), changes in call density may also indicate changes in abundance, behavior, population health, site turnover, or disturbance, which are difficult to capture using a binary detection-nondetection framework. Given a classifier score $z$ and threshold $t$, the detection rate $P(z>t)$ is related to call density by the law of total probability:
\begin{equation}\label{eqn:threshprob}
P(z>t) = P(z>t|\oplus) P(\oplus) + P(z>t|\ominus) (1 - P(\oplus)).
\end{equation}

\begin{figure}[!h]
    \centering
    \includegraphics[width=.75\textwidth]{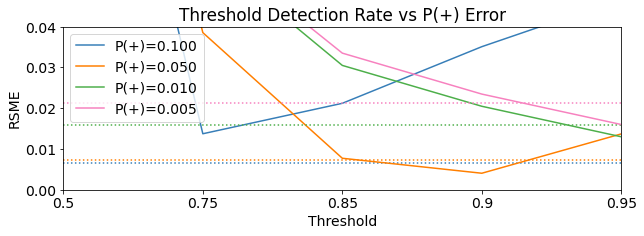}
    \caption{Root mean squared error (RMSE) between detection rates at various thresholds and true call density $P(\oplus)$, using synthetic data and a model with 0.9 ROC-AUC. Notice that the optimal threshold depends on $P(\oplus)$, which may vary across sites. Dotted lines indicate RMSE for the proposed validation scheme with 4 bins and 50 observations per bin.}
    \label{fig:thresh_quality}
\end{figure}

This simple relationship highlights an important challenge of using threshold-based detection counts as proxies for call density. $P(z>t)$ is only equal to $P(\oplus)$ when both $P(z>t|\ominus)=0$ (no false positives) and $P(z>t|\oplus)=1$ (no false negatives), which can only occur with perfect classifiers that are hard to produce. Consequently, we expect detections to over-count (due to false positives) and under-count (due to false negatives) in an unknown ratio. In Figure \ref{fig:thresh_quality}, we demonstrate a comparison between threshold detection rates and ground-truth call density using synthetic data; the optimal threshold for balancing false positives and negatives depends heavily on the true and, unfortunately, unknown call density we aim to quantify.

As we examine different subsets of the data, such as spatiotemporal slices, we expect the detection rate $P(z>t)$ to vary. Equation \ref{eqn:threshprob} demonstrates that the detection rate may vary because of changes in call density (which we hope to capture), but also in response to changes in the false-positive and false-negative rates. For example, a territorial species may use readily identifiable vocalizations near a nest but change call type or rate further away from its territory  \cite{Reid2022}, and this behavior might shift further depending on the season  \cite{Odum1955}. This would manifest as a change in site-specific $P_s(z>t|\oplus)$, independent of any change in actual call density. Likewise, the presence or absence of another species with a similar call can manifest as a change in site-specific $P_s(z>t|\ominus)$, again independent of the call rate of the target species. These examples characterize what is known as a \emph{distribution shift}. Distribution shifts are ubiquitous and underlie many difficulties in using detection counts. In bioacoustics, it is well known that thresholds selected on one dataset targeting a particular false-positive rate will not apply to new datasets because of distribution shifts \cite{Knight2018}, but shifts can easily manifest in subsets of datasets as well.

In this work, we begin developing a `threshold-free' bioacoustic analysis framework that directly estimates call density $P(\oplus)$. With the threshold discarded, Equation \ref{eqn:threshprob} becomes a statement about the full distribution of classifier scores:
\begin{equation}
P(z) = P(z|\oplus) P(\oplus) + P(z|\ominus) (1 - P(\oplus)).
\end{equation}
We propose a validation scheme, using a fixed amount of human validation effort, to approximate $P(\oplus)$.  Using a \emph{logarithmic binning} scheme, we convert the continuous classifier scores to discrete sets and validate within each bin. Logarithmic binning focuses validation effort on higher-scoring examples, which is helpful when the classifier has decent quality and the prevalence of the target species is low. 

The validation process additionally allows us to obtain a bootstrap estimate of the distribution over possible values of $P(\oplus)$ and the construction of confidence intervals. We also obtain estimates of the distributions $P(z|\oplus)$ and $P(z|\ominus)$, and the reverse conditionals $P(\oplus|z)$, $P(\ominus|z)$. The full process is summarized in Figure \ref{fig:schematic}.

We investigate the coverage and error in the estimated probability distributions using a combination of synthetic data and simulated validations on a fully annotated dataset. We explore the impact of the validation parameters on coverage and error, as well as the impact of model quality and ground-truth call density.  Of particular note, we demonstrate that the estimates produced by our validation scheme are only lightly coupled with classifier quality: We can obtain reasonable estimates (with good confidence intervals) even with moderately good classifiers, reducing the inherent need for a `perfect' classifier. 

Finally, we examine how one might use the estimated distributions produced by validation in generalized settings: For example, to estimate call density at a particular site $s$, $P_s(\oplus)$, given the distributions estimated on the level of a complete study. Computing these distributions allows us to estimate ecological parameters directly and provides a foundation for understanding and addressing the underappreciated problem of distribution shifts.

\section{Materials and Equipment}

We used three data sources for the experiments in this paper: synthetic data, a fully-annotated dataset from Pennsylvania, and a collection of PAM recordings from the Hawaiian Islands. The first two data sources were used for simulating the validation procedure and testing its coverage and error rates, as described in Sections \ref{sec:methods_error} and \ref{sec:methods_simulation}. The Hawaiian data was used to study the effects of distributional shifts in estimating site-level call rates using study-level validation in Section \ref{sec:methods_cross_site}.

\subsection{Synthetic Data}

We first assessed the validation procedure using synthetic data, where model quality and label density were controlled parameters. The synthetic dataset consists of pairs $x_j = (l_j, z_j)$ where $l_j \in \{1, 0\}$ is the ground-truth label, and $z_j$ corresponds to a model confidence score.

We may produce a `perfect' model by setting $z_i=l_i$; this combination of labels and scores has an area under the Receiver-Operator Characteristic curve score (ROC-AUC) 1.0. We can produce an all-noise model by choosing a random score $z^n_i$ for each label. We draw the random scores from a unit Gaussian with mean $0.5$. The noise model will have an ROC-AUC near 0.5.

Using these extremes we can interpolate between perfect scores and a set of noisy scores $z^n_i$ to create a model of arbitrary quality. Given a \textit{noise ratio} $r$ between 0 and 1, we set $z_i = r z^n_i + (1 - r) l_i$. After producing the scores using a given noise ratio, we can then measure the ROC-AUC of the model. Empirically, we find that the ROC-AUC decreases roughly linearly from 0.98 to 0.60 as the noise ratio varies between 0.25 and 0.75. A noise ratio of $r=0.37$ yields a model with approximately 0.9 ROC-AUC, which we use as a default value.

We may freely vary the label density by changing the proportion of positive and negative labels. We can also vary the noise level to produce an arbitrary model quality. Note that ROC-AUC is not biased by label ratio (unlike many other metrics)  \cite{bioeval2024}.

\subsection{Fully Annotated Dataset}

The Powdermill dataset consists of 6.41 hours of fully-annotated dawn chorus recordings of Eastern North American birds collected at Powdermill Nature Reserve, Rector, PA  \cite{Chron2021}. The data comprise 16,052 annotations from 48 species. The annotations allow us to compute a ground-truth call density for each species, which can be used to test the validation procedure, as described in Section \ref{sec:methods_simulation}. This dataset has particularly high label quality, as multiple expert reviewers labeled each segment, but is somewhat smaller in scope with six hours of data.

We obtain scores for each 5 second audio window in the dataset using the Google Perch bird vocalization classifier (\url{https://tfhub.dev/google/bird-vocalization-classifier/4}). The classifier has an EfficientNet-b1 convolutional architecture, and is trained on over 10k species appearing on Xeno-Canto \cite{Ghani2023}. The robust annotations allow us to exactly compute the model's ROC-AUC for each species. The model achieves a macro-averaged ROC-AUC score of 0.83 on the Powdermill dataset, as reported in the published model card.

\subsection{Hawaiian Data}

The Hawaiian dataset consists of 17.52 hours of audio collected using Song Meters (models 2, 4, or Mini, Wildlife Acoustics Inc., Maynard, MA) in 16-bit .wav format at a sampling rate of 44.1 kHz and default gain from five sites on Hawai‘i Island: Hakalau, Hāmākua, Mauna Kea, Mauna Loa, and Pu‘u Lā‘au. These recordings were compiled from past research projects and were annotated by members of the Listening Observatory for Hawaiian Ecosystems at the University of Hawai‘i at Hilo. Using Raven Pro software (Cornell Lab of Ornithology, Ithaca, NY), annotators were asked to create a selection box that captured both time and frequency around every bird call they could either acoustically or visually recognize, ignoring those that were unidentifiable at a spectrogram window size of 700 points. Annotators were allowed to combine multiple consecutive calls of the same species into one bounding box label if pauses between calls were shorter than 0.5 seconds. Recordings were then split into 5 second segments (the length of audio segments assessed by the classifier) and the number of segments that contained an annotated vocalization were tallied for each species in order to determine annotation densities $P(\oplus)$.

The majority of recordings were collected at Hakalau Forest National Wildlife Refuge on the eastern slope of Mauna Kea, totalling 11.25 hours. Hakalau is one of the largest (13,240 ha) intact, disease-free, native wet forests in the Hawaiian archipelago \cite{USFWS_2010}, and as such it is widely viewed as having the most intact and stable forest bird community remaining in Hawaiʻi. Hakalau provides habitat for eleven native Hawaiian bird species (including five federally listed endangered species), as well as many introduced bird species  \cite{Kendall2022}. The next largest contribution of data came from the high-elevation dry forests of Pu‘u Lā‘au on the southwest slope of Mauna Kea with 5.2 hours of audio. Pu‘u Lā‘au is within Ka‘ohe Game Management Area, a mixed management area open to the public for activities such as hiking and hunting, and a site with ongoing native vegetation restoration efforts intended to preserve and restore habitat for the few remaining native bird species that live there. The remaining recordings were collected in high-elevation open habitat on the southern slopes of Mauna Loa (0.55 hours), similar habitat on the eastern slopes near the summit of Mauna Kea (0.17 hours), and at a low-elevation site in Hāmākua (0.25 hours), an anthropogenically degraded habitat. The recording locations on Mauna Loa and Mauna Kea are potential nesting sites for native endangered seabirds that build burrows in lava flow crevices \cite{Antaky2019,Day2003}. The Hāmākua site has low potential to harbor native bird species, but is densely populated by introduced bird species, and was included to assess how well our computational per-site analysis would handle absent species in an acoustically active environment. 

Within this dataset we focused on three birds native to Hawai‘i Island, one species of least conservation concern, the Hawai‘i ‘Amakihi (\textit{Chlorodrepanis virens}), one vulnerable species, the ʻŌmaʻo (\textit{Myadestes obscurus}), and one federally listed endangered species, the ʻUaʻu (\textit{Pterodroma sandwichensis}) \cite{IUCN_2024}. Since the introduction of avian malaria, Hawai‘i ‘Amakihi have become uncommon below 500 m \cite{Scott_1986}, however they are one of only two Hawaiian honeycreeper species of least conservation concern. On Hawai‘i they reach the highest densities above 1,500 m in the subalpine native forests of Pu‘u Lā‘au. The ʻŌmaʻo is an endemic thrush species that inhabits montane mesic and wet forests on the windward side of Hawai‘i Island. ʻŌmaʻo are thought to be one of the more sedentary forest birds, with high site fidelity, spending the majority of their time within a 2 ha core zone \cite{Netoskie2023}, and are therefore likely to be picked up frequently on a stationary recorder. The ʻUaʻu, also known as the Hawaiian Petrel, only nests in the Hawaiian Islands where they are threatened by introduced predators \cite{Raine_2020}. ʻUaʻu dig nesting burrows on high-elevation volcanic slopes, which they mainly only visit at night, meaning they rarely share acoustic space with non-seabird species \cite{Troy2016}.

\section{Methods}\label{sec:methods}

\begin{figure}[!t]
    \centering
    \begin{minipage}[b]{0.75\textwidth}
        \centering
        \includegraphics[width=1.0\linewidth]{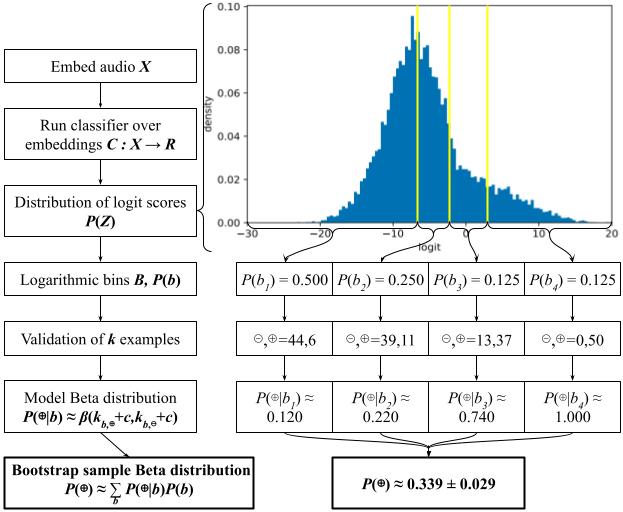}
        \label{fig:study_schematic}
    \end{minipage}
    \begin{minipage}[b]{0.75\textwidth}
        \centering
        \includegraphics[width=1.0\linewidth]{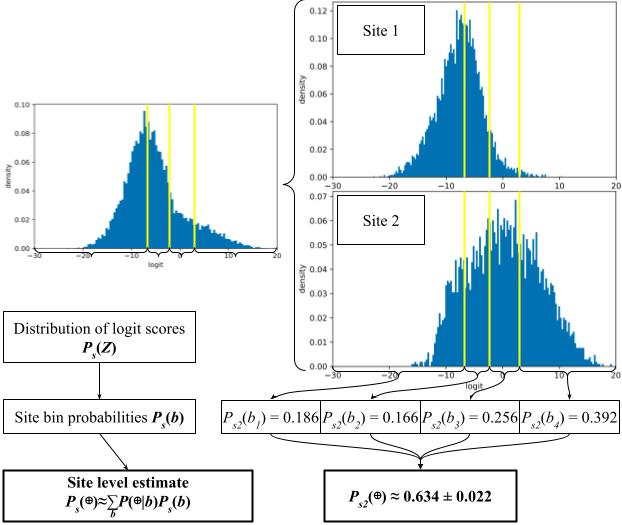}
        \label{fig:site_schematic}
    \end{minipage}  
    \caption{A schematic of our direct call density estimation method at \textbf{(A)} the study-level using our validation scheme and \textbf{(B)} the site- or covariate-level using computational Strategy 1.}
    \label{fig:schematic}
\end{figure}

\subsection{Notation}
Let $X = \{x\}$ be a large collection of audio examples, and suppose we have a trained classifier $\mathcal{C}: X \rightarrow \RR$ mapping audio examples to confidence scores for the target class. While it is not required in what follows that the confidence scores be on the logit scale, we will refer to these scores as \emph{logits}, and denote these logits with the variable $z$. Let $P(\oplus) = P(x \in \oplus)$ denote the probability that $x$ is contained in the positive set for the classifier's target class, and let $P(\ominus) = 1 - P(\oplus)$. Likewise, we will refer to distributions such as $P(\oplus|z)$, the probability that an example is in the positive class given the logit predicted by the classifier.

We may use the law of total probability to expand $P(z)$ over $P(\oplus)$:
\begin{equation}\label{eq:total_z}
P(z) = P(z | \oplus) P(\oplus) + P(z | \ominus) (1 - P(\oplus)).
\end{equation}

Or we may expand $P(\oplus)$ over $P(z)$:
\begin{equation}\label{eq:total_plus}
P(\oplus) = \int_z P(\oplus|z) P(z)
\end{equation}

We may also convert the continuous logit scores into discrete outputs by binning logits into a set of $B$ bins, $\{b\}$. We will use $b$ to refer to a generic bin, but will use $b_i$ when explicit indexing is needed. In this case, the bin probabilities $P(b)$ expand over $P(\oplus)$:
\begin{equation}\label{eq:total_b}
P(b) = P(b | \oplus) P(\oplus) + P(b | \ominus) (1 - P(\oplus)).
\end{equation}

Or we may expand $P(\oplus)$ over bins:
\begin{equation}\label{eq:total_plus_bins}
P(\oplus) = \sum_b P(\oplus|b) P(b)
\end{equation}

\subsection{Validation with Logarithmic Binning}

Because of the assumed large volume of data and low cost of running the classifier, the overall distribution $P(z)$ can be easily approximated with high accuracy. We will now describe an efficient method to estimate the distributions $P(z|\oplus)$, $P(\oplus|z)$, $P(\ominus|z)$ and $P(z|\ominus)$ with a fixed amount of human validation work, which in turn yields estimates of $P(\oplus) = \sum_z P(\oplus|z)P(z)$.

We sort the examples into $B$ logarithmic quantile bins $b_1, b_2, \ldots, b_n$ according to their logit scores, such that the lowest scoring 50\% of $X$ are assigned to $b_1$, the next lowest scoring 25\% are assigned to $b_2$, and so on, with the last bin gathering all remaining examples. With this scheme, the probability that any given $x$ falls into each bin is known (i.e., $P(x \in b_1) = 0.50$, $P(x \in b_2) = 0.25$, $P(x \in b_3) = 0.125$, etc.). 

A set of $k$ random examples are selected from within each bin for human evaluation. Each example is labeled as positive, negative, or unsure, so that for each bin $b$ we obtain counts $k_{b, \oplus}, k_{b, \ominus}, k_{b, ?}$.

The amount of human validation work required to produce these counts is fixed, given a choice of the number of bins $B$ and the number of examples to evaluate from each bin $k$.

\subsubsection{Density Estimates from Validation Outputs}

We will now demonstrate how to use the validation outputs to estimate $P(\oplus)$, $P(\oplus|b)$, and $P(b|\oplus)$ (Figure \ref{fig:schematic}A).

We model each bin with a beta distribution $P(\oplus|b) \approx \beta(k_{b, \oplus} + c, k_{b, \ominus} + c)$, where $c$ is a small constant, representing an uninformative prior for the Beta distribution. $k_{b,\oplus}$ and $k_{b,\ominus}$ can be zero so the constant $c$ is added to meet parametric constraints of the Beta distribution. Note that the when $k_{b, ?} > 0$ the total $k_{b, \oplus} + k_{b, \ominus}$ is reduced, increasing the variance of the Beta distribution.

We know $P(b)$ precisely from the logarithmic binning design. Then, using the law of total probability: 
\begin{equation}\label{eq:pplus}
P(\oplus) = \sum_b P(\oplus|b)P(b) \approx \sum_b \beta(k_{b, \oplus} + c, k_{b, \ominus} + c)P(b),
\end{equation}
and
\begin{equation}
P(b|\oplus) = \frac{P(\oplus|b)P(b)}{P(\oplus)} \approx 
\frac{\beta(k_{b, \oplus} + c, k_{b, \ominus} + c)P(b)}{\sum_d \beta(k_{d, \oplus} + c, k_{d, \ominus} + c)P(d)}.
\end{equation}

Because $P(\oplus)$ is modeled as a weighted sum of Beta distributions, we can compute its expected value as the weighted sum of the expected values of the per-bin Beta distributions.

We obtain a bootstrap distribution for $P(\oplus)$ by sampling the per-bin Beta distributions repeatedly and applying Equation \ref{eq:pplus}. We define a 90\% confidence interval for $P(\oplus)$ using the 5th and 95th quantiles of the bootstrap distribution.

\subsubsection{Estimating Model ROC-AUC from Validation Outputs}\label{sec:rocauc}
The ROC-AUC metric is a useful threshold-free indicator of model quality. In addition to its eponymous interpretation as the area under the Receiver-Operator Characteristic curve, it also has a straightforward probabilistic interpretation, as the probability that a uniformly-chosen positive example is ranked higher than a uniformly-chosen negative example \cite{bioeval2024}.

Our proposed validation procedure produces estimates of $P(b|\oplus)$, $P(b|\ominus)$. One can use these to estimate the model's ROC-AUC on the target data by summing the probability that a positive example is chosen from a higher bin than a negative example, and assuming a 50\% probability that the positive example is ranked higher when they come from the same bin:
\begin{equation}\label{eq:rocauc}
ROC-AUC \approx \sum_{i>j} P(b_i|\oplus) P(b_j|\ominus) + \frac{1}{2} \sum_{i} P(b_i|\oplus) P(b_i|\ominus).
\end{equation}

\subsubsection{Evaluating Coverage and Error of Density Estimates}\label{sec:methods_error}

Our validation procedure responds to effectively five parameters, which can be categorized in three types: First, the Beta distribution prior is a \textit{general hyper-parameter}. Second, we have \textit{user parameters}: the number of bins and the number of validations per bin, which determine the total human effort required, and impact the quality of estimates. Finally, we have \textit{extrinsic parameters}, out of control of the user: The actual prevalence of the target signal in the dataset and the quality of the classifier.

To measure the \textit{coverage} of our estimate of $P(\oplus)$, we checked whether the ground-truth density was within the 90\% confidence interval of the bootstrap distribution for $P(\oplus)$ around 90\% of the time, measured over a large number of trials. We found a value of Beta prior which provides good coverage for both synthetic and fully annotated data, across a wide range of ground-truth densities.

We measured the \textit{precision} of the validation estimate by tracking the root mean squared error (RMSE) of the expected value of $P(\oplus)$ over a large number of trials. We examined the response in error to choice of binning strategy (uniform or logarithmic), changes in the number of bins, number of validated observations per bin, and model quality.

Computing the coverage and precision of validation requires access to the ground-truth density, which we had for both the synthetic data and Powdermill fully-annotated data.

\subsubsection{Simulation Experiments with Annotated Dataset}\label{sec:methods_simulation}

To assess the ability of our validation procedure to reliably estimate $P(\oplus)$, we leveraged the fully-annotated Powdermill dataset and a pre-trained bird species classifier. For this experiment, we simulated the entire validation process by leveraging the existing human annotations. For each example we selected for validation, we checked whether the example overlapped a human annotation, and said it was a positive example if there was any overlap. We computed the expected value of $P(\oplus)$ from the validation data, as described above, and measured the RMSE from the ground-truth value. We obtained a ground-truth value $P^{GT}(\oplus)$ from the annotations by counting the proportion of segments which overlapped annotations for the target species. Finally, we checked whether $P^{GT}(\oplus)$ landed in the 90\% confidence interval produced by the bootstrap estimate.

\subsection{Distribution Shifts: Site and Covariate Estimates}\label{sec:methods_strats}

It is common for a PAM project to span many microphones, placed at various \emph{sites}. We refer to any geotemporal subset of the observations $X$ as a \emph{site} $X_s$. Likewise, we may have a covariate $V$ and can refer to a subset selected by covariate value $v$ as $X_v$.

When restricting from the study-level set of observations to a site or covariate subset, we expect to observe distribution shifts: In fact, observing and explaining such distribution shifts is a major reason to engage in bioacoustic monitoring in the first place. Let us consider these distribution shifts in the context of Equation \ref{eq:pplus} (Figure \ref{fig:schematic}B), and compare to the situation when using a threshold-based binary classification scheme.

We are likely most interested in measuring the change in $P_s(\oplus)$ between sites and by comparison to the study-level $P(\oplus)$. Thanks to copious observations, we can observe any changes in $P_s(z)$ easily, which we expect will correspond to changes in the relative abundance or activity of the target species. And indeed, our equations tell us that this $P_s(z)$ decomposes as:
\begin{equation}
P_s(z) = P_s(z | \oplus) P_s(\oplus) + P_s(z | \ominus) (1 - P_s(\oplus)).
\end{equation}

While any of the three distributions on the right-hand side of this equation might shift, it is $P_s(\oplus)$ which we wish to isolate. All of $P_s(\oplus)$, $P_s(z | \oplus)$, and $P_s(z | \ominus)$ are unknown, which means that we need to introduce either new assumptions or additional data to estimate $P_s(\oplus)$.
We can obtain estimates for $P_s(\oplus)$ in a few different ways. 

\textbf{Strategy 0}: First, we can add more data. Applying our validation procedure on data from each site will certainly yield per-site estimates of $P_s(\oplus)$, though we expect that this validation will be onerous if the number of sites and/or target species is high.

Alternatively, we can substitute in study-level estimates of the component distributions to allow us to isolate $P_s(\oplus)$.

\textbf{Strategy 1}: We might assume that $P_s(\oplus|z) = P(\oplus|z)$ to leverage knowledge gained from validation at the study-level. Then we have:
\begin{equation}
P_s(\oplus) = \int_z P_s(\oplus|z)P(z) dz \approx \int_z P(\oplus|z)P(z) dz.
\end{equation}

Or, using our logarithmic binning:
\begin{equation}
P_s(\oplus) \approx \sum_b P(\oplus|b)P_s(b).
\end{equation}

This is very straightforward to compute, but the distribution shift between $P(\oplus|z)$ and $P_s(\oplus|z)$ may be problematic. Notice that $P_s(\oplus|z) \propto P_s(z|\oplus) P_s(\oplus)$, and thus depends vitally on the parameter we want to estimate! In particular, if the site is unoccupied by the target species, then $P_s(\oplus)=0$, so that $P_s(\oplus|b)=0$ as well.

This strategy is analogous to the application of a binary classifier to a new subset of the data: counting threshold detections typically assumes that the true-positive rate with respect to a given threshold is fixed as we look at different subsets of data.

\textbf{Strategy 2}: Another approach is to assume that $P_s(b|\oplus) = P(b|\oplus)$: The distribution of logits for positive examples is roughly the same across sites. This seems a far more reasonable assumption: The vocalizations of the target species, wherever it is present, are similar. However, it turns out that this is not enough to solve directly for an estimate of $P_s(\oplus)$.

To obtain our estimate, we utilize the decomposition over binned logits, additionally substituting the study-level distribution $P(b|\ominus)$:
\begin{equation}
P_s(b) \approx P(b|\oplus) P_s(\oplus) + P(b|\ominus) (1 - P_s(\oplus)).
\end{equation}
We obtain a distribution over logit bins for any value $q\in [0, 1]$, specifying an arbitrary mixture of the positive and negative logit distributions:
\begin{equation}
Q_q(b) = P(b|\oplus) q + P(b|\ominus) (1 - q).
\end{equation}
Then for any choice of $q$, we may compute the KL-Divergence $KL(P_s(b) || Q_q(b))$, which is the cost of substituting $Q_q(b)$ for the ground-truth distribution $P_s(b)$. We then set:
\begin{equation}
P_s(\oplus) \approx \text{argmin}_q KL(P_s(b)||Q(q))
\end{equation}

The downside of this strategy is that we are vulnerable to shifts in both the positive and negative distributions, relative to the study-level.

\textbf{Strategy 3}: Because Strategies 1 and 2 use quite separate heuristics for obtaining estimates of $P(\oplus)$, they can be combined as an ensemble estimate by taking their geometric mean. 

In Section \ref{sec:cross_site}, we compare all four strategies (per-site validation, substitution of $P_s(\oplus|b)$, substitution of $P_s(b|\oplus)$, and ensemble estimation) on the Hawaiian PAM data.

\subsubsection{Distribution Shift in Real-world Data}\label{sec:methods_cross_site}

Finally, we used real-world PAM deployment data from Hawaiʻi to compare varying approaches to estimating site-level densities. We compared the results of site-specific validation to substitution of study-level distributions $P(\oplus|z)$ or $P(z|\oplus)$. 

Feature embeddings were extracted from the recordings using the pre-trained Google Perch model. We then trained a linear classifier over the pre-computed embeddings using examples from 7 native bird species, and 6 common non-native bird species, with variable numbers of training samples (Table \ref{tab:ha_training}), following the procedure in \cite{Ghani2023}. None of the training examples were sourced from the PAM recordings used in this study. The classifier was then run over the embedded PAM data and a logit score was generated for each 5 second segment within the dataset for each of the three study species.

\begin{table}[H]
\centering
\caption{Hawaiian classifier training data} 
\begin{tabular}{lc}
\toprule
\textbf{Class} & \textbf{Train Examples} \\ \midrule
ʻAkēʻakē & 509 \\ 
ʻApapane & 3284 \\ 
Erckels Francolin & 56 \\ 
Eurasian Skylark & 233 \\ 
Hawaiʻi ʻAmakihi & 1158 \\ 
Hawaiʻi ʻElepaio & 1096 \\ 
Iʻiwi & 1756 \\ 
Northern Cardinal & 95 \\ 
ʻŌmaʻo & 2046 \\ 
Red-billed Leiothrix & 138 \\ 
ʻUaʻu & 775 \\ 
Warbling White-eye & 120 \\ 
Yellow-fronted Canary & 96 \\ 
Other/Unknown & 372 \\ \bottomrule
\end{tabular}
\label{tab:ha_training}
\end{table} 

We then applied the validation scheme, using 4 bins and 50 examples per bin for Hawaiʻi ʻAmakihi and ʻŌmaʻo for a total of 200 examples for each species. Because of their low density at the study-level, 200 examples per bin were validated for ʻUaʻu for a total of 800 examples. Using Equation \ref{eq:pplus} $P(\oplus)$ was estimated for each species. 

Site-level estimates $P_s(\oplus)$ were then computationally estimated for each study species using the methods described above in Section \ref{sec:methods_strats}. Manual site-level validation was then performed for Hawaiʻi ʻAmakihi for both the Hakalau and Pu‘u Lā‘au datasets to generate validated site-level estimates $P_{sv}(\oplus)$. All manual validations were performed by an acoustic specialist trained for Hawaiian bird species (AKN).

\section{Results}\label{ref:results}

\subsection{Validation Coverage}

\begin{figure}[!t]
    \centering
    \begin{minipage}[b]{0.49\textwidth}
      \centering
      \includegraphics[width=1.0\linewidth]{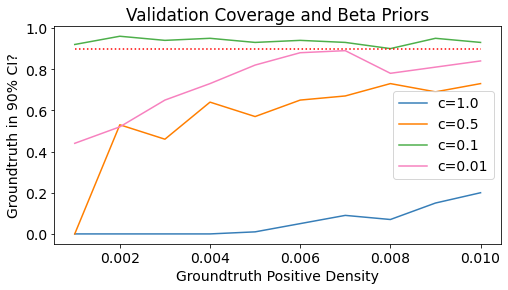}
    \end{minipage}
    \hfill
    \begin{minipage}[b]{0.49\textwidth}
      \centering
      \includegraphics[width=1.0\linewidth]{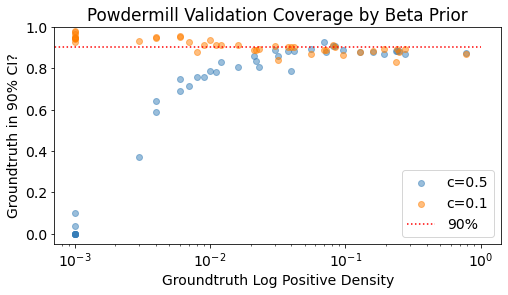}
    \end{minipage}
    \caption{For different uninformative Beta distribution priors, we run simulation studies to find how often the ground-truth prediction of $P(\oplus)$ is in the predicted 90\% confidence interval. The value $c=0.1$ has better coverage at low call density to the Jeffrey's prior or the uniform prior, both on \textbf{(A)} synthetic data and in \textbf{(B)} Powdermill simulations. All experiments use 4 bins and 50 observations per bin.}
    \label{fig:beta_reliability}
\end{figure}

We first investigated the coverage of the predicted $P(\oplus)$. We found that coverage was typically good when the ground-truth density $P(\oplus)$ was above 0.1, and depended on the choice of Beta distribution prior at lower densities. As shown in Figure \ref{fig:beta_reliability}, we had good coverage at low density with the prior $c=0.1$. For the synthetic data, we used the default noise value corresponding to a classifier with ROC-AUC 0.9. For the Powdermill data, classifier quality varied widely by species, demonstrating good coverage with the 0.1 prior regardless of classifier quality. We fixed $c=0.1$ in all subsequent experiments.

\subsection{Validation Error}

\begin{figure}[!h]
    \centering
    \includegraphics[width=.75\textwidth]{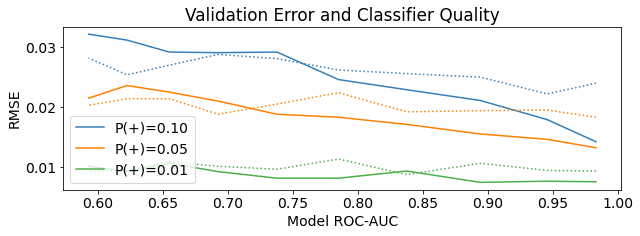}
    \caption{Root mean squared error (RMSE) of the predicted $P(\oplus)$ in synthetic data, demonstrating that error steadily decreases as model quality improves, and above 0.75 ROC-AUC, and that logarithmic binning gives lower error above 0.75 ROC-AUC. RSME's are means over 50 trials. Solid lines report results with logarithmic binning, and dotted lines report results with standard, evenly spaced bins.}
    \label{fig:dense_quality}
\end{figure}

We examine the relationship between classifier quality and RMSE of the predicted $P(\oplus)$ on synthetic data, varying the classifier quality between 0.58 and 0.98, and at three different ground-truth densities (Figure \ref{fig:dense_quality}). We found that improving classifier quality generally decreased the RMSE. We also compared our logarithmic binning scheme to a uniform binning approach, and found that logarithmic binning generally gave a lower RMSE once the classifier ROC-AUC was greater than 0.75. Note that we would expect no improvement for classifiers with ROC-AUC 0.5, since the higher logits are equally likely to be positive or negative examples.

\begin{figure}[!h]
    \centering
    \begin{minipage}[b]{0.49\textwidth}
      \centering
      \includegraphics[width=1.0\linewidth]{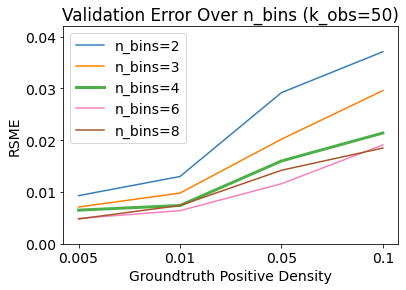}
    \end{minipage}
    \begin{minipage}[b]{0.49\textwidth}
      \centering
      \includegraphics[width=1.0\linewidth]{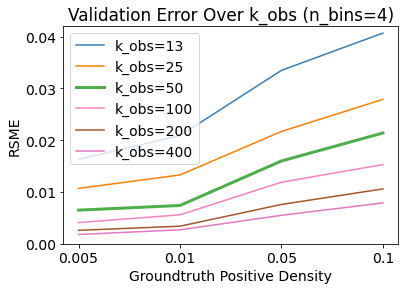}
    \end{minipage}
    \caption{The two user parameters for validation are the number of bins ($n_{bins}$) and the number of validation examples per bin ($k_{obs}$). Here we demonstrate, using the synthetic data harness, variation in the precision of $P(\oplus)$ as we vary the \textbf{(A)} number of bins and \textbf{(B)} observations per bin: Adding more data (more bins, or more observations) generally leads to lower root mean squared error (RMSE). For this model, at 0.9 ROC-AUC, error saturates at 4-6 bins, but decreases steadily as more observations per bin are added.}
    \label{fig:val_precision}
\end{figure}

We examined the change in density prediction error on synthetic data as we varied the number of bins and number of observations per bin (Figure \ref{fig:val_precision}). In the logarithmic binning scheme, increasing the number of bins only splits observations at the top end of the logit distribution: For high quality classifiers, the highest bins may already be purely positive, so that increasing the number of bins adds no new information. Thus, we observed that eventually there was no improvement in error as more bins were added.

On the other hand, increasing the number of observations per bin steadily decreased the prediction error, as we would expect: The per-bin Beta distributions become narrower and more precise, which translates into more precise predictions of $P(\oplus)$.

Thus, we found that for a reasonably good classifier, four bins is likely sufficient, and additional effort is better spent by increasing the number of observations per bin, rather than further increasing the number of bins.

\begin{figure}[!h]
    \centering
    \includegraphics[width=.75\textwidth]{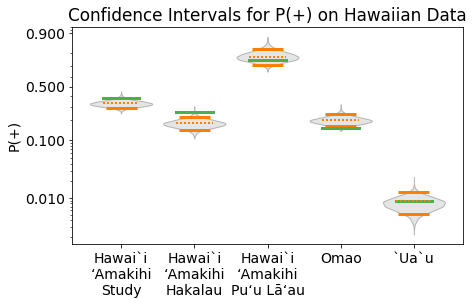}
    \caption{Confidence intervals for $P(\oplus)$ on Hawaiian PAM data. Green lines are the annotation density, which approximate ground-truth. Solid orange lines give bounds of the 90\% confidence interval, and dotted orange lines give the expected value of $P(\oplus)$. The distributions of bootstrap-sampled $P(\oplus)$ values are in grey. Note that there is only a 59\% chance that all five annotation density values would fall inside accurate 90\% CIs.}
    \label{fig:hawaii_cis}
\end{figure}

\subsection{Cross-Site Prediction}\label{sec:cross_site}
Manual annotation performed by the Listening Observatory for Hawaiian Ecosystems found Hawai‘i ‘Amakihi were present at Hakalau with an annotation density of 0.241 and were the most acoustically active passerine species at Pu‘u Lā‘au with an annotation density of 0.748, for an overall study-level annotation density of 0.380. ʻŌmaʻo were only present at Hakalau where their site-level annotation density was 0.240 resulting in a density of 0.155 at the study-level. ʻUaʻu vocalizations were present at both Mauna Kea and Mauna Loa, with annotation densities of 0.192 and 0.256, respectively, though because there were few recordings from these locations, their study-level annotation density was only 0.009. None of these species were present at the Hāmākua site.

The results of all strategies for cross-site prediction, as well as call densities produced by manual validation (Figure \ref{fig:hawaii_cis}), are provided in Table \ref{tab:ha_site_transfer}. No validation was performed on sites known to be unoccupied.

\begin{table}[H]
\centering
\caption{Annotation density (Ann) and site-level predictions of $P(\oplus)$ via validation (Val), substitution of study-level $P(\oplus|b)$ (Strat 1) and substitution of study-level $P(b|\oplus)$ (Strat 2), and the geometric mean of Strategies 1 and 2 (Strat 3). ROC-AUC is estimated from validation data, using Equation \ref{eq:rocauc}.}
\begin{tabular}{lcl ccccc}
\textbf{Species} & \textbf{ROC-AUC} & \textbf{Site} & \textbf{Ann} & \textbf{Val} &  \textbf{Strat 1} & \textbf{Strat 2}  & \textbf{Strat 3}
\\[0.5ex]\hline &&&& \\[-1.5ex]
Hawaiʻi     & 0.84 & Study-level      & 0.380 & 0.333 & -     & -     & -     \\
ʻAmakihi    &      & Hakalau     & 0.241 & 0.180 & 0.202 & 0.039 & 0.089 \\  
            &      & Pu‘u Lā‘au  & 0.748 & 0.770 & 0.640 & 0.991 & 0.797 \\ 
            &      & Hāmākua     & 0.000 & U     & 0.123 & 0.000 & 0.000 \\
            &      & Mauna Kea   & 0.000 & U     & 0.263 & 0.000 & 0.000  \\
            &      & Mauna Loa   & 0.000 & U     & 0.205 & 0.000 & 0.000 
\\[0.5ex]\hline &&&& \\[-1.5ex]
ʻŌmaʻo      & 0.78 & Study-level      & 0.155 & 0.200 & -     & -     & - \\
            &      & Hakalau     & 0.240 & -     & 0.255 & 0.346 & 0.297 \\
            &      & Pu‘u Lā‘au  & 0.000 & U     & 0.100 & 0.000 & 0.000  \\
            &      & Hāmākua     & 0.000 & U     & 0.098 & 0.000 & 0.000  \\
            &      & Mauna Kea   & 0.000 & U     & 0.090 & 0.000 & 0.000  \\
            &      & Mauna Loa   & 0.000 & U     & 0.088 & 0.000 & 0.000
\\[0.5ex]\hline &&&&&& \\[-1.5ex]
ʻUaʻu       & 0.88 & Study-level & 0.009 & 0.009 & -     & -     & - \\ 
            &      & Hakalau     & 0.000 & U     & 0.000 & 0.000 & 0.000 \\
            &      & Pu‘u Lā‘au  & 0.000 & U     & 0.020 & 0.190 & 0.062 \\
            &      & Hāmākua     & 0.000 & U     & 0.017 & 0.142 & 0.049 \\
            &      & Mauna Kea   & 0.192 & -     & 0.068 & 0.972 & 0.258 \\
            &      & Mauna Loa   & 0.256 & -     & 0.056 & 0.772 & 0.208 \\ 
\end{tabular}
\label{tab:ha_site_transfer}
\end{table}

\begin{table}[H]
\centering
\caption{Distribution shifts between study-level estimates for Hawaiʻi ʻAmakihi and site-level distributions at Puʻu Lāʻau. For Hawaiʻi ʻAmakihi the study-level unconditional distribution of confidence scores over the bins was heavier in the lower bins relative to the site-specific distribution of confidence scores. For Puʻu Lāʻau the decomposition of the distribution of confidence scores over bins into positive and negative components revealed a large shift in the positive distribution relative to the site-level validation. Starred distributions are produced by site-level validation.}
\begin{tabular}{cc|cccc}
            & \textbf{Distribution}  & $b_1$ & $b_2$ & $b_3$ & $b_4$ \\[0.5ex]
 All logits       & $P(b)$                 & 0.50  & 0.25  & 0.12  & 0.12 \\
 ~                & $P_s(b)$               & 0.19  & 0.17  & 0.26  & 0.39 
 \\[0.5ex]\hline &&&& \\[-1.5ex]
 $\oplus$ logits  & $P(b|\oplus)$          & 0.18  & 0.17  & 0.28  & 0.38 \\
                  & $P_s(b|\oplus)^*$      & 0.03  & 0.16  & 0.31  & 0.50 
\\[0.5ex]\hline &&&& \\[-1.5ex]
 $\ominus$ logits & $P(b|\ominus)$         & 0.66  & 0.29  & 0.05  & 0.00 \\
                  & $P_s(b|\ominus)^*$     & 0.76  & 0.19  & 0.05  & 0.00 \\
\end{tabular}
\label{tab:dist_shift}
\end{table}

Overall, Strategy 1 struggled when a site was unoccupied: Weight in low bins is still assigned to the target species, as expected. On the other hand, Strategy 2 generally predicted non-occupied sites correctly: no weight in the high bins implies that there is no contribution from the target species.

Meanwhile, we found that Strategy 2 often overestimated call density when a site was occupied. Examining the actual study- and site-level distributions, we found cases (such as the Hawaiʻi ʻAmakihi at Pu‘u Lā‘au, detailed in Table \ref{tab:dist_shift}, and the ʻUaʻu at Mauna Kea) where the study-level $P(b|\oplus)$ was extremely similar to the site-level $P_s(b)$. In such cases, Strategy 2 selects a $P_s(\oplus)$ very close to 1.

Strategy 3 did a surprisingly good job of balancing the strengths and weaknesses of Strategies 1 and 2. Particularly in the case of unoccupied sites, Strategy 2 often correctly predicted $P_s(\oplus)=0$, so that the geometric mean was zero. And, at least in this study, we found that for occupied sites Strategy 1 tended to underestimate while Strategy 2 overestimated, leading to improved estimates in the geometric mean.

\section{Discussion}\label{ref:discussion}
Our method for directly estimating call densities in bioacoustic data from machine learning classifier outputs yielded promising results and could advance the field of PAM by expediting analysis and providing a framework for formal ecological hypothesis testing. This approach is less dependent on highly performant and consistent classifiers because it utilizes the entire distribution of model outputs to estimate study-level and site-specific distributions, which makes it less reliant on consistent decision boundaries around arbitrary threshold levels. We also found the distributions over logarithmic bins helpful in identifying and describing distribution shifts, a pervasive but underappreciated problem in bioacoustic analyses.   

\subsection{Validation Quality}
Both the simulated data and Powdermill validation simulations offer ground-truth values on which we tested the quality of our validation scheme. We found that a single choice of Beta prior gives strong coverage across a wide variety of ground-truth densities and classifier qualities. The Powdermill validation simulations build confidence by including a scenario with real data distributions and a wide variety of call densities and classifier qualities.

While we found that error decreased with increasing model quality, we also found that adding additional validation effort reduced error. This provides a path to improvement for practitioners with access to a pre-trained classifier without further machine learning effort.

\subsection{Handling Distribution Shifts in Real-World Data}   
From the Hawaiian PAM dataset, our method produced study-level call density estimates with 90\% confidence intervals that contained the manual annotation densities obtained by trained technicians for Hawaiʻi ʻAmakihi and ʻUaʻu but produced a slight overestimate for ʻŌmaʻo. Site-level validations (Strategy 0) similarly achieved estimates close to annotation density values, as would be expected with additional user effort. The manual annotation procedure was not explicitly designed for this study, and some errors may have been introduced in our derivation of annotation densities. For instance, individual vocalizations may have been split into separate 5-second segments. If either portion of the split call was too short to be identified, it would be counted as a negative during validation but marked as a positive annotation for both segments. 

This PAM dataset also represents a high level of heterogeneity between sites, ranging from the acoustically active Hakalau to the windswept Mauna Loa, which is nearly devoid of any species vocalizations. This inherent heterogeneity in acoustic characteristics led us to expect the large distribution shifts observed in our analysis. Each of our focal species revealed strengths and challenges to our computational strategies for estimating call densities at the site-level via extrapolation from the study-level distributions of $P(\oplus)$ and $P(\ominus)$, and $P_s(b)$. 

\subsubsection{Insights from Hawai`i `Amakihi - Cross-Site Comparison}
Not only were Hawai‘i ‘Amakihi at much lower densities at Hakalau than Pu‘u Lā‘au, Pu‘u Lā‘au also has less competition for acoustic space, and therefore the distributions of $P(z|\oplus)$ and $P(z|\ominus)$ were significantly different for ‘amakihi at these two sites. Because of this, differing computational strategies performed best for predicting call densities each site. At Pu‘u Lā‘au, where the distribution of $P_s(b)$ closely tracked $P(b|\oplus)$, Strategy 3 outperformed the others, balancing out the underestimate from Strategy 1 and overestimate from Strategy 2. However, at Hakalau, where $P_s(\oplus| z)$ tracked the study-level $P(\oplus|z)$ Strategy 1 made the closest estimate. It is worth noting that all strategies showed distinctly higher acoustic activity levels at Pu‘u Lā‘au, the more active site.

Table \ref{tab:dist_shift} shows the exact distribution shifts at Pu‘u Lā‘au. Shifts in the negative logits $P_s(b|\ominus)$ and positive logits $P_s(b|\oplus)$ almost exactly canceled out, making $P_s(b)$ closely resemble $P(b|\oplus)$, leading to over-prediction in Strategy 2. The distribution shifts in Table \ref{tab:dist_shift} are also interesting to consider from a threshold-detection perspective. If the boundary between $b_3$ and $b_4$ were used as a threshold, we would see a significantly increased true positive rate at Pu‘u Lā‘au. If one only counted detections at the site-level assuming the same true positive rate as at the study-level, one would infer a too-high estimate of activity change.

Leveraging knowledge about species-habitat associations and vocalization behavior to inform the stratification process of the validation procedure may improve the call density estimates. For example, we could validate the logit distributions across stratified covariates and leverage site-specific covariate values to develop conditional site-level distribution estimates. We hope to explore such approaches in later work.

\subsubsection{Insights from `Oma`o - Cross-Species Confusions and Classifier Quality}
When a classifier was trained solely on ʻŌmaʻo, a large portion of logits fell into the higher-level bins at Mauna Kea and Mauna Loa (data not shown). This may be because both ʻŌmaʻo and ʻUaʻu, as well as another seabird species present at these sites, often use low-frequency vocalizations. While our validation protocol yielded similar study-level estimates to those obtained using a classifier trained on all species in Table \ref{tab:ha_training}, the site-specific call density 90\% confidence intervals did not encompass 0 at sites containing seabird vocalizations. Adding non-class training examples from acoustically similar species shifted ʻŌmaʻo logits out of those top bins and yielded the estimates shown in Table \ref{tab:ha_site_transfer}. 

This emphasized three things: First, our proposed validation procedure lessens our dependency on developing high-performance classifiers, as reliable study-level estimates can be made even when only one species is included in the model. Second, including acoustically similar species can boost model performance, shifting a large portion of the positive examples into the top bins, which subsequently increases the \textit{precision} of call density estimates by lowering RMSE and improving site-level estimation. Last, observing a strong shift from study-level distributions may indicate when modifications to study design or additional validation efforts are necessary for robust site-level estimates.

\subsubsection{Insights from `Ua`u - Heavy Top Bin}
The ʻUaʻu was the only species for which Strategy 2 (and thus Strategy 3) incorrectly predicted presence in unoccupied sites. At the study-level, ʻUaʻu has a very low overall prevalence. With four bins of 200 examples each, all validated observations of the ʻUaʻu were obtained in the top bin. Additionally, \emph{nearly all} logits at Mauna Kea (97.5\%), and to a lesser extent at Mauna Loa (79.9\%), landed in the top bin. We believe that the heavy wind on the otherwise quiet mountain tops was a useful discriminative feature for the target class (a nocturnal petrel), leading to relatively high logits for all examples at these sites and from other sites containing wind with few vocalizations. However, the classifier still ranks windy examples with the target species higher than windy examples without the target species. In essence, all of the interesting site-level variation is subsumed in the top bin.

This problem could be addressed in a few ways. First, adding more bins at the top should split the positive-windy examples from the negative-windy examples. Second, the study-level validations could be restricted to the mountain environments (Mauna Kea and Mauna Loa), so the windy logits are distributed more evenly over the bins. We would also expect less extreme shifts between the study-level and site-level logit distributions, bolstering the substitution assumptions in Strategies 1 and 2. Finally, the too-broad highest bin suggests a role for a continuous distribution estimate (such as a Kernel Density Estimate) instead of a binned estimate. We leave exploring these options further for later work.

\subsection{Conclusions and Future Research}
Ecological inference increasingly relies on predictive modeling, especially with the widespread adoption of sensor-based sampling methods that rely on computational algorithms. A key challenge lies in navigating sets of predictions and making informed decisions amidst uncertainty. The process of classification is a decision process \cite{Spiegelhalter1986} that requires disentangling the processes of predictive modeling and decision-making under uncertainty, as well as appropriate tools, such as scoring measures to assess predictive performance \cite{Steyerberg2010, Resin2023}. However, threshold-based approaches often conflate these steps, limiting flexibility and potentially leading to sub-optimal decisions (in this case, classifications). Our work takes an important step forward by considering sets of probabilities and surfacing relevant parameters for optimizable utility functions based on available resources, such as lab capacity. Our findings demonstrated that performance increased with more reviewed audio clips, indicating that biacoustic programs can leverage this structured, data-driven approach to allocate their resources adaptively.

The protocol we have proposed here for directly assessing call densities in bioacoustic data has significant applications in the field of avian conservation. Our approach's relatively low time cost facilitates analysis of PAM data within actionable timeframes, which can boost the utility of monitoring in informing wildlife management decisions \cite{Nichols2006, Gibb2018}. For example, the State of Hawaiʻi Department of Land and Natural Resources and U.S. Fish and Wildlife Service are currently taking actions to mitigate avian malaria mortality in forest birds native to Hawaiʻi \cite{Warner_1968} by suppressing the population of its mosquito vector (\textit{Culex quinquefasciatus)} \cite{KauaiIIT}. For malaria-sensitive species, changes in juvenile call densities, a reasonable indicator of fledgling survival, estimated using our approach could be used to assess the efficacy of mosquito control efforts. Further,  our methods could provide a standardized approach for analyzing past PAM data to establish historical baselines and assess changes to biodiversity over time with fine spatiotemporal resolution.

While the work described here has great potential, it serves as a preliminary tool, and we foresee multiple potential routes to improvement. First and foremost, future work should focus on improving covariate-level call density estimates. One potential way to do so may be to validate samples along strata or gradients relevant to the ecological or detection process of interest (e.g., along an elevational gradient) instead of validating to bins of the study-level distribution. This would mitigate the distribution shift issues encountered in our study. In addition to distribution shifts along environmental or temporal gradients, shifts in vocalization behavior could also lead to domain shifts. Future work could investigate the effect of separating species-level classifiers into call-type classifiers (i.e., separate classes for `songs,' `contact calls,' `begging'), which could improve classifier score calibration at the study-level and thereby improve covariate-level call density estimates. Call-type classifiers would have the additional benefit of aiding in modeling behavior and ecology.

\printbibliography

\section*{Conflict of Interest Statement}
Author Tom Denton is employed by Google. The remaining authors declare that the research was conducted in the absence of any commercial or financial relationships that could be construed as a potential conflict of interest.

\section*{Author Contributions}
TD and AKN conceived the study design and assisted in writing the manuscript. TD developed the methods proposed in this article, and performed all experiments on simulated data and the fully annotated Powdermill dataset. AKN provided suggestions for the methods described here, and performed all experiments on the Hawaiian PAM dataset. MJW provided suggestions for study design and interpretation, and assisted in writing the manuscript. PJH provided support for the study and assisted in revising the manuscript. All authors contributed to the article and approved the submitted version.

\section*{Funding}
This work was partially funded by a grant awarded to PJH and AKN by the United States Geological Survey FY23 National and Regional Climate Adaptation Science Center Program (\#G23AC00641).

\section*{Acknowledgments}
We want to thank the members of the Listening Observatory for Hawaiian Ecosystems who assisted in the manual annotation of recordings, including Noah Hunt, Saxony Charlot, Nikolai Braedt, and, in particular, lab manager Ann Tanimoto-Johnson. Lauren Harrell in Google Research provided helpful feedback throughout this work. Members of the Google Perch team (Bart van Merri\"{e}nboer, Jenny Hamer, Vincent Dumoulin, Eleni Triantafillou, Rob Laber) contributed to creating the Perch embedding model and active learning methods used to produce the classifiers used in this paper.

\section*{Data Availability Statement}
Our Hawaiian PAM dataset and classifier scores, as well as annotation information for the three study species, can be found on Zenodo at https://doi.org/10.5281/zenodo.10581530. The fully annotated Powdermill dataset assembled by Chronister et al. that was used in this study is available at  https://doi.org/10.1002/ecy.3329.

\section*{Figure captions}
Figure 1. Root mean squared error (RMSE) between detection rates at various thresholds and true call density $P(\oplus)$, using synthetic data and a model with 0.9 ROC-AUC. Notice that the optimal threshold depends on $P(\oplus)$, which may vary across sites. Dotted lines indicate RMSE for the proposed validation scheme with 4 bins and 50 observations per bin.

Figure 2. \textbf{(A)} the study-level using our validation scheme and \textbf{(B)} the site- or covariate-level using computational Strategy 1.

Figure 3. For different uninformative Beta distribution priors, we run simulation studies to find how often the ground-truth prediction of $P(\oplus)$ is in the predicted 90\% confidence interval. The value $c=0.1$ has better coverage at low call density to the Jeffrey's prior or the uniform prior, both on \textbf{(A)} synthetic data and in \textbf{(B)} Powdermill simulations. All experiments use 4 bins and 50 observations per bin.

Figure 4. Root mean squared error (RMSE) of the predicted $P(\oplus)$ in synthetic data, demonstrating that error steadily decreases as model quality improves, and above 0.75 ROC-AUC, and that logarithmic binning gives lower error above 0.75 ROC-AUC. RSME's are means over 50 trials. Solid lines report results with logarithmic binning, and dotted lines report results with standard, evenly spaced bins.

Figure 5. The two user parameters for validation are the number of bins ($n_{bins}$) and the number of validation examples per bin ($k_{obs}$). Here we demonstrate, using the synthetic data harness, variation in the precision of $P(\oplus)$ as we vary the \textbf{(A)} number of bins and \textbf{(B)} observations per bin: Adding more data (more bins, or more observations) generally leads to lower root mean squared error (RMSE). For this model, at 0.9 ROC-AUC, error saturates at 4-6 bins, but decreases steadily as more observations per bin are added.

Figure 6. Confidence intervals for $P(\oplus)$ on Hawaiian PAM data. Green lines are the annotation density, which approximate ground-truth. Solid orange lines give bounds of the 90\% confidence interval, and dotted orange lines give the expected value of $P(\oplus)$. The distributions of bootstrap-sampled $P(\oplus)$ values are in grey. Note that there is only a 59\% chance that all five annotation density values would fall inside accurate 90\% CIs.

\end{document}